# Advanced Clustering Techniques for Speech Signal Enhancement: A Review and Meta-analysis of Fuzzy C-Means, K-Means, and Kernel Fuzzy C-Means Methods

**Abdulhady Abas Abdullah[1], Aram Mahmood Ahmed[1,3], Tarik Rashid[1,3], Hadi Veisi[2]**

[1] **Artificial Intelligence and Innovation Centre, University of Kurdistan Hewler, Erbil, Iraq.**

[2] **School of Intelligent Systems College of Interdisciplinary Science and Technologies University of Tehran**

[3] **Computer Science and Engineering Department, University of Kurdistan Hewler, Erbil, KR, Iraq.**

## Abstract:

Speech signal processing is a cornerstone of modern communication technologies, tasked with improving the clarity and comprehensibility of audio data in noisy environments. The primary challenge in this field is the effective separation and recognition of speech from background noise, crucial for applications ranging from voice-activated assistants to automated transcription services. The quality of speech recognition directly impacts user experience and accessibility in technology-driven communication. This review paper explores advanced clustering techniques, particularly focusing on the Kernel Fuzzy C-Means (KFCM) method, to address these challenges. Our findings indicate that KFCM, compared to traditional methods like K-Means (KM) and Fuzzy C-Means (FCM), provides superior performance in handling non-linear and non-stationary noise conditions in speech signals. The most notable outcome of this review is the adaptability of KFCM to various noisy environments, making it a robust choice for speech enhancement applications. Additionally, the paper identifies gaps in current methodologies, such as the need for more dynamic. Clustering algorithms that can adapt in real time to changing noise conditions without compromising speech recognition quality. Key contributions include a detailed comparative analysis of current clustering algorithms and suggestions for further integrating hybrid models that combine KFCM with neural networks to enhance speech recognition accuracy. Through this review, we advocate for a shift towards more sophisticated, adaptive clustering techniques that can significantly improve speech enhancement and pave the way for more resilient speech processing systems.

**Keywords:** speech Enhancement, noise reduction, clustering algorithms, K-Means (KM), Fuzzy C-Means (FCM), Kernel Fuzzy C-Means (KFCM)

## 1. Introduction

Clustering is an unsupervised machine-learning technique that groups data points based on their similarities. In this method, both real objects (like a number gathered by a sensor) or non-real objects (e.g., a document) can be used. Some of the most used clustering techniques are K-Means, Fuzzy C-Means and Kernel Fuzzy C-Means out of which diversified usages can be identified in fields such as data mining, pattern recognition etc. (Zhang and Zhao, 2024). Although K-Means is more popular due to its simplicity and quick performance in many applications, FCM has the measures of membership degrees which provide more flexibility and accuracy necessary while dealing with datasets having overlapping clusters. It is an extension of FCM with added kernel methods to manage non-separable incidences; to excel it is widely applied in various complicated subject fields comprises of speech signal processing than KFCM (Vani et al., 2019). Thus, the uses of these clustering techniques are diverse, including data analysis, applications in biological data analysis or biotechnology, machine learning, image processing and others. That is why, due to the nonlinearity and non-stationary nature of the speech data, the methods mentioned above are appropriate, as they act rather correctly managing uncertainty and providing not only binary results, but rather spectrum of the memberships, that can improve the performance of the model in comparison with the basic techniques (Singh and Singh., 2024). In the present review, we compare KM, FCM, and KFCM and presented the importance of these methods for analysis and improvement of speech signals. Unlike previous works, this review takes some features into account such as, Euclidean distance algorithmic complexity, blurring effects, stopping criteria, execution times and recognition precision (Kumar et al., 2020). The study aims at determining the applicability of these techniques in addressing noisy additive and speech signals using different types of datasets which are homogeneous and heterogeneous. The article especially highlights the advantage of Kernel Fuzzy C-Means (KFCM) clause in handling aggregated and noisy signals, as highlighted by Aradnia et al. (2022). This comprehensive review is called for by publication trends indicating the annual output and interest in these clustering techniques especially FCM and KFCM is rising steadily. An analysis of major academic databases reveals a significant annual increase in publications related to these methods, indicating a sustained interest and ongoing advancements in their applications (Source: Name of Database, Library Journal Database Analysis, 2024). The main purpose of this review is to provide the academic community with novel information's on the methodological developments and various applications of KM, FCM, and KFCM. It aims at filling existing gaps by presenting recent findings and discussing the newest areas of applications in modern fields including image processing, agriculture, data mining, and others (Czarnowski and Jędrzejowicz, 2018; Khanlari and Ehsanian, 2017). This work not only helps researchers to improve their research in speech signal processing but also encourages them to explore new directions further in respect of new application areas and improvements on these methodologies in other areas where uncertainty and data fluctuation exist. Figure 1 shows a rising trend in publications for Kernel Fuzzy C-Means and Fuzzy C-Means in speech enhancement from 2017 to 2024.

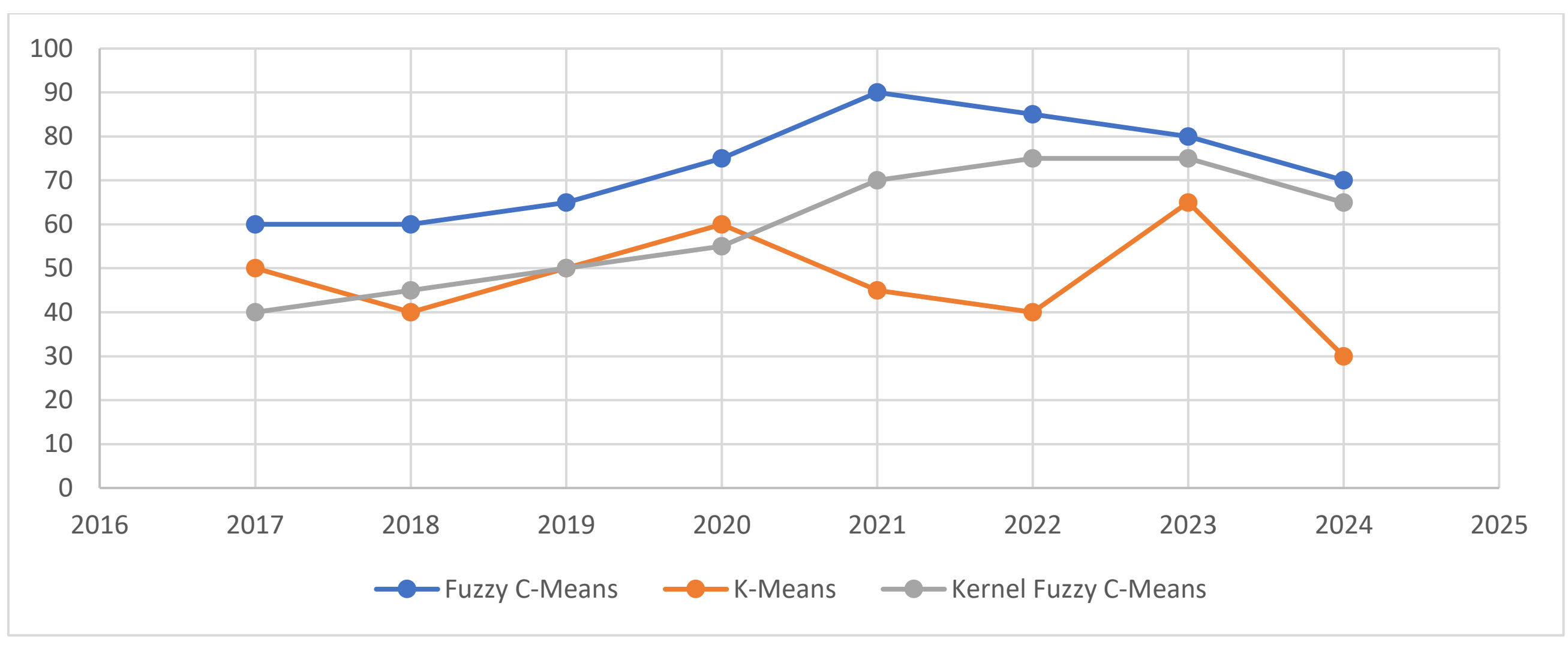

Figure 1: Several Yearly Published Articles Cited The Fuzzy C-Means, K-Means, and Kernel Fuzzy C-Means Methods Used in Speech Enhancement and Effective Noise Reduction in Speech Signals Since 2017.

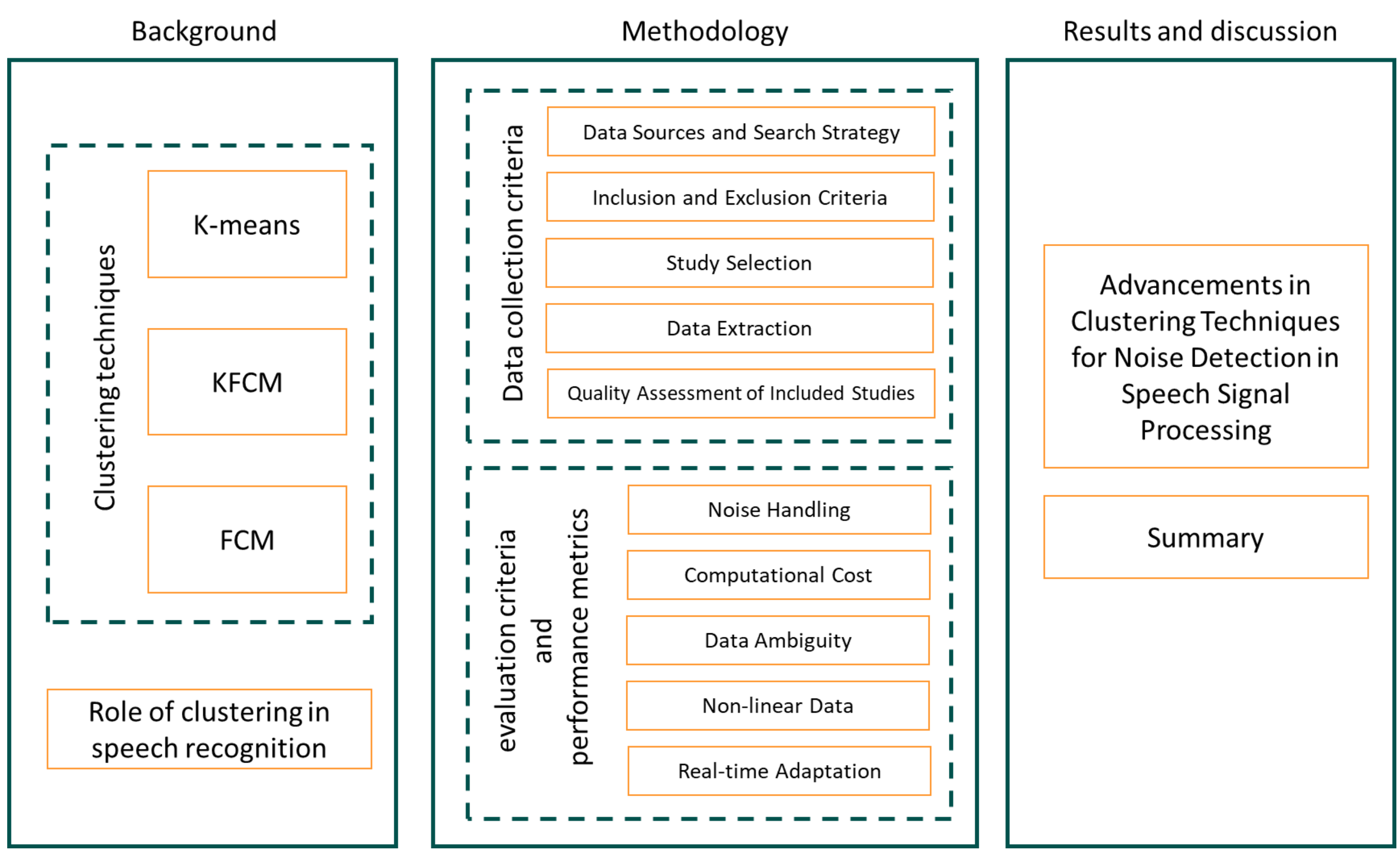

Figure 2: Organizational Structure of the Article

The structure of this paper, as shown in figure 2, is organized as follows: Section 2 delves into the theoretical foundations of fuzzy clustering techniques, with a particular focus on FCM and KFCM, outlining their core algorithms and parameters. It also discusses the role of clustering in speech recognition. Section 3 presents the used methodology to conduct this study including data collection criteria as well as evaluation criteria and performance metrics. Section 4 explores advancements in clustering techniques for noise detection and enhancing speech signals. Finally, Section 5 concludes the paper with a summary of key findings, contributions to the field, and suggestions for future research directions.

## 2. Background

Speech signal processing has become of significant significance for the contemporary need of digital communication covering the areas of telecommunication, automated transcription, and voice activated systems (Abdullah et al., 2024). The main difficulty is speech recognition from noisy background which is crucial in enhancing the quality of communication and the effectiveness of speech-based systems (Abdullah et al.,2024). The process of speech signal processing contains several steps, which are summarized in figure 3:

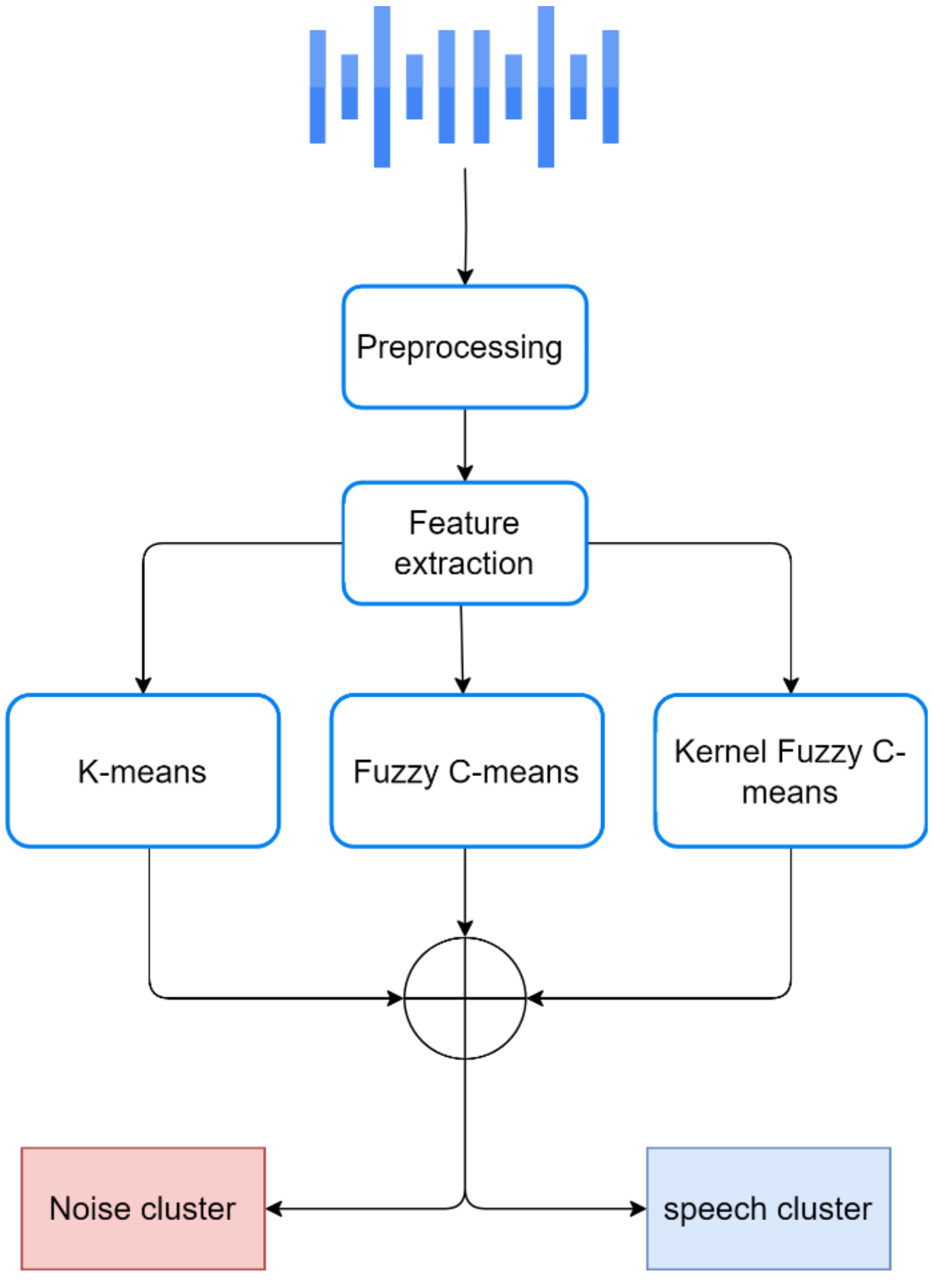

Figure 3: Overview of the clustering process in speech signal processing.

• **Data Input:** The first stream of data is in the form of speech information of a given session.

• **Preprocessing:** The data is first preprocessed to reduce noise and to normalize the data.

• **Feature Extraction:** The preprocessed data is used to derive the features that are naturally grouped in clusters.

• **Clustering:** Various methods of clustering (K-Means, Fuzzy C-Means, and K-Partitional Fuzzy C-Means) are used to cluster the data points.

• **Clustered Output:** These clusters can then be used for further processing or even for analysis.

Each is important in Preprocessing and Postprocessing the speech to ensure that good quality information is passed in the communication and to ensure reliability in noisy environment. The techniques discussed, that includes K-Means, FCM, and KFCM, are different ways of parting data into clusters where the fuzzy clustering methods FCM and KFCM, include more complex features for handling uncertainty to enhance clustering (Salman and Alomary., 2024).

## 2.1 Clustering Techniques

Clustering is used to identify close relationships between data. In the other hand, fuzzy logic can identify to which degree, each data belongs to their respective groups. Therefore, clustering methods are divided into two categories: soft-clustering and hard-clustering (Hosseini et al.,2024). soft clustering uses the concept of fuzzy membership functions such as triangular or Gaussian to group the data so that each point of data becomes a member of two or more groups with different membership values (Lin and Busso., 2024). Hard clustering, on the other hand, assigns each data point to exactly one cluster, without any ambiguity about its membership. In the literature, there are a large number of clustering algorithms. However, this study focuses on Fuzzy C-Means, K-Means, and Kernel Fuzzy C-Means Methods merely (Liu et al., 2024).

### 2.1.1 K-Means Clustering Technique

K-means clustering is one of the most fundamental and widely used clustering algorithms due to its simplicity and efficiency. It is designed to partition a dataset into distinct clusters by minimizing the within-cluster variance. The algorithm starts with an initial set of centroids, which can be selected randomly or through heuristic methods such as the K-Means algorithm that improves initialization (Ahmed and Hassan, 2023). During each iteration, K-Means assigns each data point to the nearest centroid, forming clusters based on Euclidean distance. Centroids are then updated as the average of all the data points that are assigned to specific clusters after the assignment process. This process happens iteratively until the convergence, which is distinguished by the small change in the centroid's location or the change in the cluster's assignment.

Figure 4 A. illustrates the step-by-step process of the K-Means algorithm:

**Initialization:** Randomly choose k initial centroids or by some heuristic method.
**Assignment:** The next step consists in labelling each data point with the centroid which is nearest to it according to the Euclidean distance.
**Update:** About the centroid update rule, new centroids are determined as the average coordinate of all data points belonging to a particular cluster.

**Convergence Check:** Perform the assignment and update steps again and again until the current and the new centroids are almost the same.

K-Means is a popular algorithm; however, it has some drawbacks. It becomes a factor of the number of clusters k which must be determined by the user in advance, in case the right one is unknown. Thus, it can be noted that the algorithm is very dependent on the initial positioning of the centroids, and this is likely to result into different clusters on different uses of the same data set. In addition, it makes certain assumptions like spherical clusters and equal cluster size which might not be very useful for a dataset having clusters of varying shapes and densities (Abdullah and Veisi., 2022). Therefore, there are several important implications of K-Means; though it is one of the efficient algorithms particularly when dealing with big data and generally gives good solutions, it has some restrictions on the type of data it is fitting for, and the parameters of the algorithm must be chosen accurately. For speech enhancement, centroid drift and cluster switching can occur as noise characteristics change frame-to-frame, degrading separation. Moreover, Euclidean distances in raw features may not reflect perceptual similarity. Practical deployments often need sliding-window or online K-Means, feature normalization, and temporal constraints to stabilize assignments in dynamic conditions.

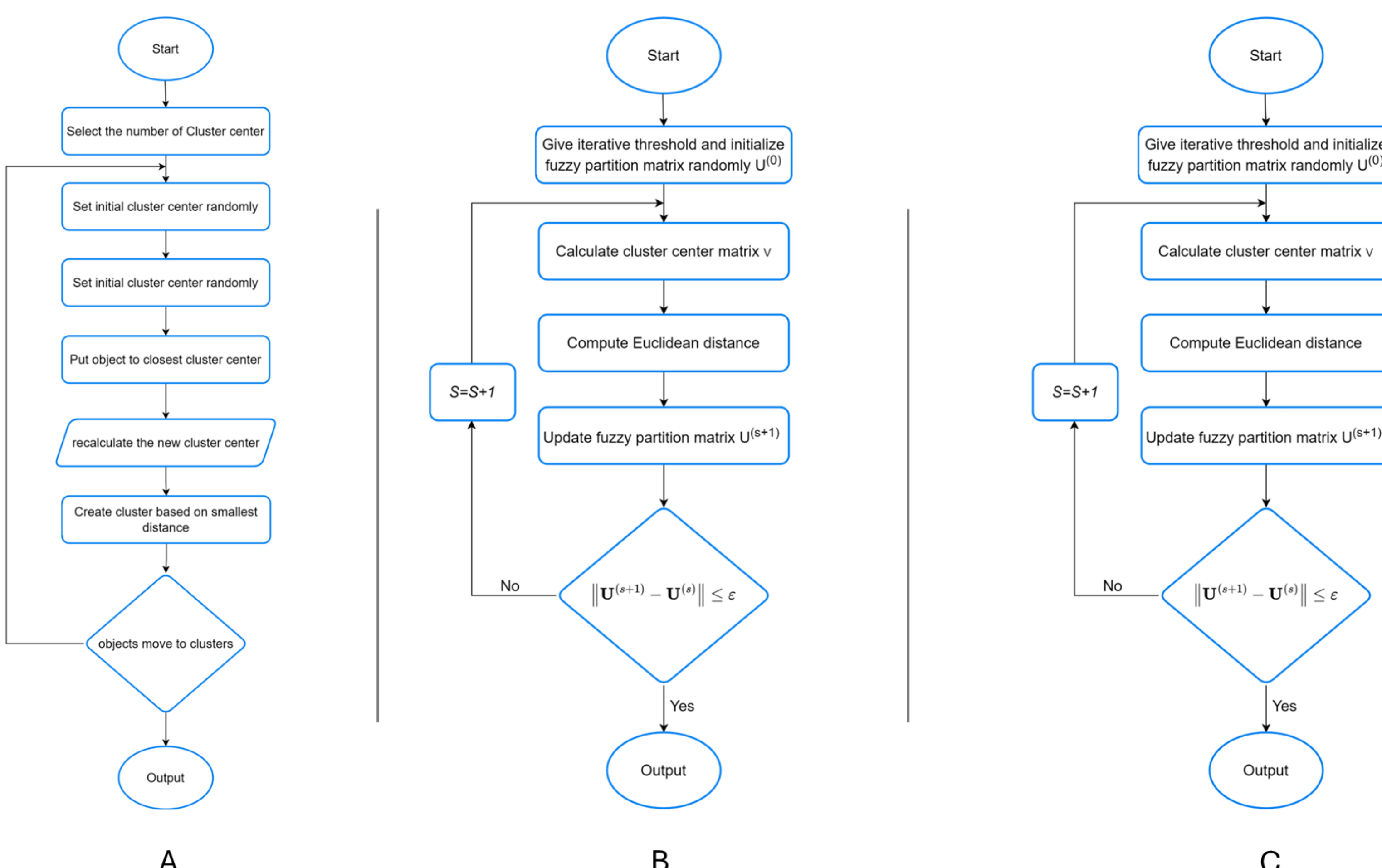

Figure 4 A: K-Means Clustering Algorithm Process Flowchart, B: fuzzy c-means clustering algorithm Flowchart, C: kernel fuzzy c-mean clustering flowchart.

### 2.1.2 FCM Clustering Technique:

Compared with the category III, Fuzzy C-Means (FCM) is more flexible where every single data point can belong to multiple clusters at the same time with some levels of membership. This method is especially used when the data points cannot be easily classified or when there is blurring between clusters. FCM works by Endowing each data point with membership values, which describe how much the data point belongs to each cluster. These membership values are then used to compute the cluster centers as weighted means of the data points where the weight is the respective membership value (Yu et al., 2024).

Figure 4 B. depicts the process of FCM:

**1. Initialization:** Assign the first memberships and the initial positions of the cluster centers.

**2. Update Membership:** Assign fuzzy membership values for each data point about their distances to the cluster centers and on the fuzziness parameter $m$.

**3. Update Centers:** Find new cluster centers by averaging the new membership values concerning all data points belonging to each cluster.

**4. Convergence Check:** Perform the update steps until convergence, that is, the membership values and the cluster centers do not change significantly.

FCM has been proved to be a useful tool for most of the applications due to its flexibility to allow overlapping of clusters such as image segmentation where the pixels fall under more than one region and with different intensity levels. Soft partitioning appears to give more about the structure of the data and the membership values offer more fine-grained segmentation patterns (Dhal et al., 2024). However, FCM is dependent on the parameters for example the fuzziness parameter which was represented as $m$ determines the degree of overlapping between the clusters. Thirdly, there is a limitation of being computationally costly for large set because the membership matrix is gradually modified through iteration. However, FCM is still a valuable technique for clustering data that by their nature contain uncertainty and are blurred between clusters (Moosavi et al., 2023). In dynamic audio scenes, FCM's soft memberships help at speech/noise boundaries, but online updates and careful selection of $m$ are needed to prevent over-smoothing speech transients or lagging behind abrupt noise onsets. Computation must be budgeted for real-time pipelines.

### 2.1.3 KFCM Clustering Technique

KFCM is an improved method of FCM because it takes advantage of the kernel methods, which makes it easier to solve problems of data that cannot be well classified and clustered using the traditional methods. The idea of KFCM is that kernel functions should spread the points into a higher dimensional space where there may be much better cluster separation. This is a management transformation that allows KFCM to preserve detailed structures of data that cannot be easily separated in the original space (Singh and Srivastava, 2024).

Figure 4 C. illustrates the steps involved in KFCM:

**1. Kernel Mapping:** Use kernel function to map the lower or higher original data into a higher-dimensional feature space.

**2. Initialization:** Assign the first value to the members and centers of the clusters in the transformed space.

**3. Update Membership:** Assign the weights of membership to the level of the distances in the feature space of data points from the centers of different clusters.

**4. Update Centers:** Update new importance values of clusters as the average of data points weighted based on membership values.

**5. Convergence Check:** Perform the update steps until the membership values, and the cluster centers patterns, are stabilized.

Due to kernel methods, KFCM can capture complex dependencies and structures of data in a direct way, which is essential for datasets with a high degree of complexity. They often use it in fields like pattern analysis and in the biomedical field due to the high level of complexity. Nonetheless, the selection of the kernel function and its parameters substantively affects the convergence of KFCM, and the time complexity raises with a high value of the feature space. These challenges can nevertheless be explained by the fact that KFCM offers a strong substrate for confronting complicated clustering issues and improving clustering quality in severe conditions (Shang et al., 2019). In speech enhancement, kernels can sharpen the separation between speech harmonics and noise components, especially under adverse, rapidly varying interference. However, kernel selection and hyperparameters must be constrained to avoid overfitting and to keep streaming latency acceptable.

Importantly, these benefits come with non-trivial computational and memory costs: dense kernelization increases per-iteration work (many kernel evaluations per frame) and, if a Gram matrix is cached, can induce $O(n^2)$ memory is untenable for streaming or embedded settings. In frame-synchronous pipelines (e.g., 10 ms hops), the added kernel computations can raise end-to-end latency and reduce throughput relative to K-Means/FCM, and overly sharp kernels (large $\gamma$ in RBF) it may slow convergence or overfit transient noise. To balance precision and efficiency in real-time, we recommend approximate kernelization (Nyström landmarks or Random Fourier Features to an explicit $D$-dimensional map with FCM in $\mathbb{R}^D$ ), sliding-window operation with warm-starts, early-stopping (e.g., max 20-40 iterations or $\Delta < \varepsilon$ for 3 steps), and two-stage gating that invokes KFCM only on uncertain frames ( $u_{\max} < \tau$ These measures retain much of KFCM's non-linear accuracy while keeping latency close to FCM. [Added] In summary, KFCM is preferable when robustness near speech/noise boundaries at low SNR is critical, whereas K-Means (latency-first) or FCM (ambiguity-tolerant but lighter-weight) remain better defaults under tight CPU/RAM or power budgets.

## 2.3 Role of Clustering Techniques in Speech Enhancement and Noise Reduction

Speech processing incorporates a broad set of approaches related to the analysis, generation and modification of speech. These techniques are applied in many areas such as in speech recognition, speaker recognition and speech, options (Batra et al., 2024). Speech recognition is a process of transforming spoken language to text, in real application such as voice operated devices and transcription (Abdullah et al., 2024). The applications of speaker identification and verification are based on the voice pattern to assert or recognize identity for security and services needed (Muhamad et al., 2024). Another important field within speech processing is the one of speech enhancement, which constitutes improving the quality and acoustical clarity of the speech signal, by reducing the input of different kinds of distortions such as the background noise (Mraihi and Ben Messaoud., 2024). A speech enhancement system is essential in many environments, for example, mobile communication, as well as hearing aids where clear speech is important (Patil and Pandhare., 2025). This may incorporate use of complex mathematical models meant to filter the signals of speech from the rest of the noise to ensure the final speech is still intelligible (Upadhyay., 2024). Other applications of SP are speech synthesis where computer generated voices are used to convert written text into voice (Ahmad and Rashid., 2024). This technology is used in assistive devices and virtual agents and automated customer care, where natural sounding and contextually suitable speech is required (Huq et al., 2024). Linguistic and acoustic characteristics of ASR and speech synthesis algorithms should be complemented to achieve high quality synthetic speech (Tan et al., 2024). Clustering is an important technique when it comes to noise reduction within the FP realm since data that form a cluster are grouped together in order to separate speech from noise. Noise reduction in clustering algorithms seeks to differentiate between speech and the different types of noises that can be applied different procedures (Xu and Wunsch, 2023). Due to efficient data sorting in terms of an audio feature collection, these algorithms can enhance the quality of speech, focusing on noise components removal without losing the values of speech quality (Lee et al., 2024). A simple yet effective method for noise reduction employs K-Means clustering for the proper categorization and identification of audio makers into disparate clusters that define the speech and noise characteristics (Saha et al., 2024). This method is appreciated from the computational point of view and its simplicity; for this reason, it applies to situations where speech and noise are sufficiently separated (Wang and Watanabe., 2024). Nevertheless, K-Means algorithm depends on the number of clusters set in advance and may face problems with overlapping data (Ikotun et al., 2023). K-Means algorithm on the other hand has some limitations when applied to the DSP domain because of the overlapping nature of the speech and noise components This is because K-Means centers all the data points of the input space and allows only one cluster at a time for every data point It is for this reason that Fuzzy C-Means (FCM): a development of the K-Means approach permits data points to belong to more than This approach is useful in noisy situations where there is a closeness of the dominant speech to noise (Hashemi et al., 2023). However, FCM can be more computationally demanding, and there is necessary to tune parameters in order to get the best solution. Kernel Fuzzy C-Means (KFCM) is therefore an improvement of FCM as it integrates kernel methods with the objective of dealing with non-separable data (Singh and Srivastava, 2024). As data are mapped into a higher dimensionality, KFCM can learn the transformation between speech and noise in a complex fashion that is especially beneficial in adverse noise conditions (Hu et al., 2023). However, KFCM is sensitive to kernel function and its computational process takes more time (Tang et al., 2023). Thus, the choice and parameterization of suitable clustering techniques are critical for noise elimination, in general. Both these clustering techniques provide unique benefits and drawbacks that affect the system's performance in various noise environments (Miraftabzadeh et al., 2023).

## 2.4 Real-World Applications

Clustering underpins practical speech front-ends across products where labels are scarce or delayed. In telecommunications, K-Means/FCM drive robust VAD and pre-codec denoising with tight latency budgets; soft memberships reduce boundary errors in babble and street noise. In voice assistants (smart speakers, mobile), frame- and TF-domain clustering supports wake-word pre-screening, far-field enhancement, and prompt segmentation; FCM improves stability around speech/noise transitions, while KFCM (with Nyström/RFF) cuts false rejects under highly non-stationary scenes (TV, kitchen) without breaking the ~100 ms wake budget (Miraftabzadeh et al., 2023). Hearing aids/hearables rely on ultra-low-power clustering for scene classification (speech-in-noise, wind, music) and artifact-aware noise reduction; soft labels reduce "pumping," and lightweight kernel approximations yield intelligibility gains in rapid acoustic shifts. In call centers/conferencing, clustering aids speaker change detection, double-talk/echo control, and channel quality monitoring; FCM/KFCM stabilizes diarization when overlaps and room changes occur, improving downstream ASR segmentation. Automotive hands-free systems cluster spatial/spectral cues to separate driver speech from engine/road noise; kernelized variants help with cabin non-linearities when approximated for embedded DSP constraints. Across these domains, selection follows constraints: K-Means for strict latency/energy, FCM for ambiguity tolerance, and KFCM (approx.) when non-linear, non-stationary noise dominates and accuracy gains outweigh compute, aligning with objective measures (PESQ/STOI/SI-SDR, WER/CER) used in deployment pipelines (Singh and Srivastava., 2024).

## 3. Methodology

It thus describes how clustering algorithms for speech signal enhancement and noise reduction; K-Means, Fuzzy C-Means (FCM), and Kernel Fuzzy C-Means (KFCM) can be assessed. These algorithms were chosen because they differ in their performance on the speech data and more importantly on the noisy data.

## 3.1 Data Collection Criteria

### 3.1.1 Data Sources and Search Strategy

This review empirically analyzes and integrates studies on advanced clustering algorithms used in the process of enhancing speech signals and reducing noise, with emphasis on the K-Means, FCM, and KFCM approaches. The identified studies are reviewed according to the PRISMA guidelines, which aimed to conduct a systematic and bias-free analysis. An extensive bibliographic search was carried out and these electronic databases were used: IEEE Xplore, PubMed, Scopus, Google Scholar. The research focused on studies published between 2017 and 2024. To find relevant literature for this study, the following specific keywords and keyword strings have been used: 'Advanced Clustering Techniques for Speech Signal Enhancement and Noise Reduction,' 'K-Means,' 'Fuzzy C-Means,' 'Kernel Fuzzy C-Means,' 'Speech Processing,' 'Speech Signal Enhancement,' 'Noise Reduction,' and 'Clustering Algorithms.' To make sure that relevant and high-quality papers could be included in the database, the criteria for inclusion and exclusion were used. An initial search of articles was performed, and we obtained a large list of related papers, which were subsequently filtered using title, abstract and full-text reviews. A data extraction form was developed prior the review and was applied to collect specific information from each study regarding characteristics, clustering methodologies, speech processing applications and findings. The search strategy and review process were structured as outlined in Table 1 below:

Table 1: Steps in the Data Sources and Search Strategy Process

| Step | Description |
|---|---|
| Initial Search | Retrieval of articles using specified keywords across selected databases. |
| Duplicate Removal | Elimination of duplicate entries from the initial pool of articles. |
| Title and Abstract Screening | Preliminary screening based on relevance to the topic. |
| Full-Text Assessment | Detailed evaluation of remaining articles to ensure they meet inclusion criteria. |
| Data Extraction | Systematic extraction of relevant data using a predefined form. |

In accordance with PRISMA guidelines, this activity proceeded systematically and was designed to provide an inclusive overview of recent developments in the field of advanced clustering approaches speech and noise elimination.

### 3.1.2 Inclusion and Exclusion Criteria

As necessary to maintain a relevance and quality of the works revisited in this systematic review on advanced clustering techniques for speech signal enhancement and noise reduction, inclusion and exclusion criteria for the present article were defined as follows: These criteria were intended to perform an adequate 'squeeze' on the studies, to include only the most relevant and of the highest quality available. The criteria are summarized in Table 2 below:

**Inclusion Criteria:**

1. **Publication Date:** The inclusion of studies published within the years 2017- 2024 were used to capture the recent innovations in clustering approaches for speech processing.
2. **Type of Publication:** To increase the reliability of the included studies, only journal articles and conference papers which are peer-reviewed were included.
3. **Focus on Clustering Techniques:** Originally only articles dedicated only to techniques like K-Means, Fuzzy C-Means, and Kernel Fuzzy C-Means for speech signal enhancement and noise reduction were considered for the purpose of this study. This criterion helped keep the review on the specific use of these techniques in speech processing.
4. **Empirical Results and Methodology:** Only the articles that described the findings in which at least one empirical result was provided, methodologies shared, or theoretical literature concerning clustering techniques advanced were included. This ensured that the review includes practical usage and theoretical development.

**Exclusion Criteria:**

1. **Language:** Only articles published in English were considered within the framework of the review because it is much easier to work with such material and to interpret the results for the reviewers.
2. **Primary Focus:** Literatures that were not positioned mainly on speech signal enhancement or noise reduction no matter whether they utilized clustering approaches had been ruled out. This helped to keep the review highly relevant to specific applications throughout the process.
3. **Type of Article:** Books, reviews, commentaries, opinions, editorials, correspondence, and other non-empirical articles were not included in the review because the purpose of the review involved understanding cutting-edge original research contributions in the literature**.**

Table 2**:** Inclusion and Exclusion Criteria Summary

| Criteria Type | Criterion |
|---|---|
| Inclusion | Publication Date: 2017-2024 |
| | Type of Publication: Peer-reviewed journal articles and conference papers |
| | Focus on Clustering Techniques: K-Means, FCM, KFCM in speech processing |
| | Empirical Results and Methodology |
| Exclusion | Language: Non-English publications |
| | Primary Focus: Studies not related to speech signal enhancement or noise reduction |
| | Type of Article: Review articles, opinion pieces, editorials, and non-research articles |

In pursuing the target objectives, it was necessary to use these elaborated stringent inclusion and exclusion criteria to ensure the accumulation of only the high-quality and relevant studies, which would help to widen the understanding of advanced clustering techniques for the enhancement of the speech signal and to reduce the noise. Such systematic approach made it possible to include only the studies that richen the field and contribute to a more valid review of literature.

### 3.1.3 Study Selection

The articles in this review of advanced clustering techniques for speech signal enhancement and noise reduction: K-Means, Fuzzy C-Means (FCM), and Kernel Fuzzy C-Means (KFCM) were selected following a proper research strategy in order to involve only the most relevant and valuable articles. An initial search of electronic databases yielded a total number of 1100 articles. As always, some of these sources were duplicated, and after filtering them out, 743 articles were obtained. Initially, only the articles sourced were examined about relevance based on their titles and their abstracts to influence the omission of articles not speaking of clustering techniques for speech processing. After this first step, we removed 800 articles, and from the remaining 128 articles, we proceeded to the next step. Of the 128 articles, all underwent full text screening using prespecified inclusion and exclusion criteria. The articles reviewed were assessed in terms of the articles match with the research focus, the quality of the studies and their impact on the existing literature. Due to this approach, further 300 articles were screened out as they failed to meet standards required or they lacked empirical and theoretical content. After following this process, the present review includes 78 pertinent studies. The conclusions of these studies will help to draw a general picture of the state of development of cluster-based methods for speech signal and noise filtering and analyze the prospects for their further improvement and use.

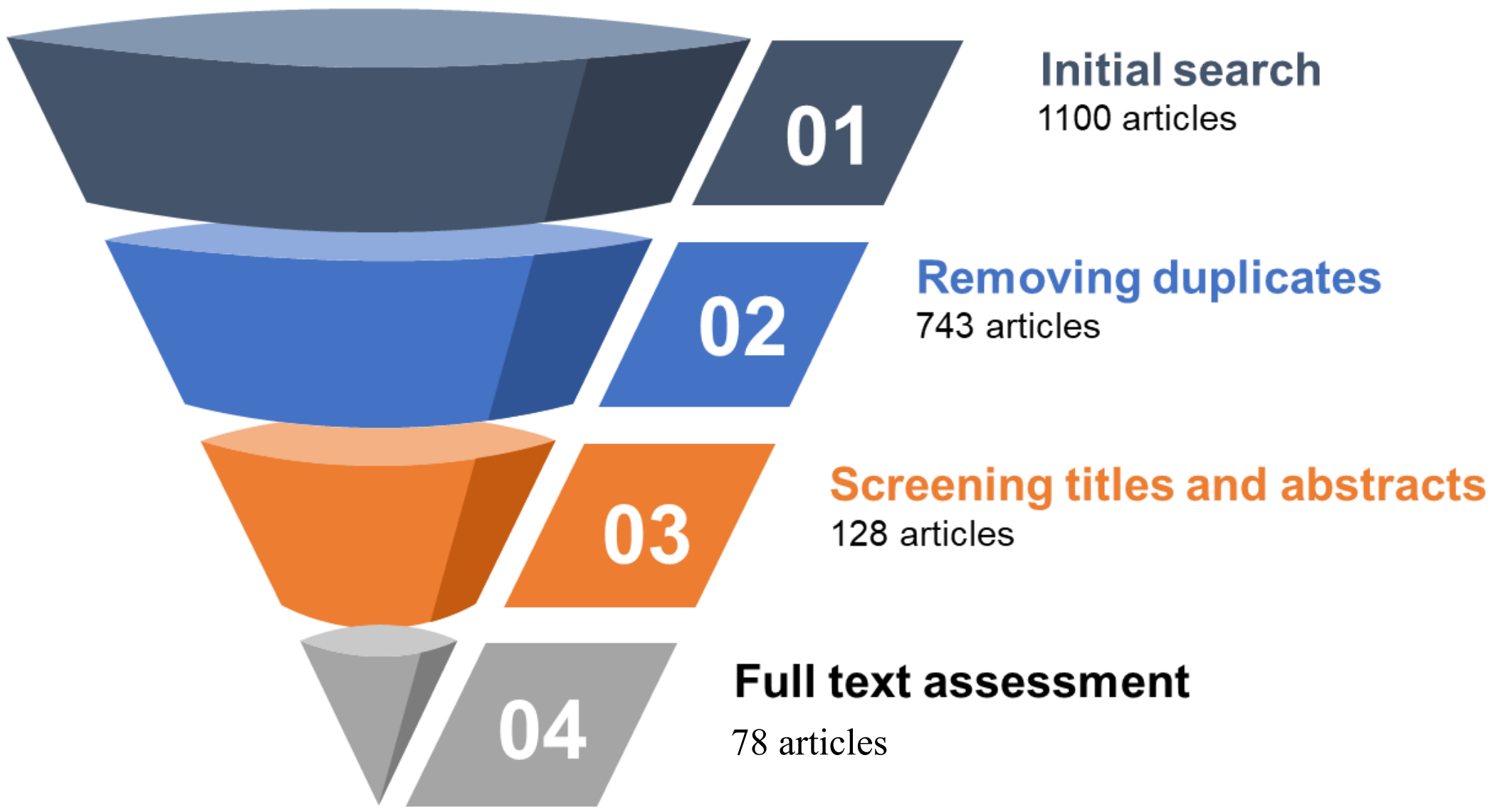

Figure 5 Study Selection Summary

Figure 5 explains the number of articles included at each stage of the review process and therefore exhibits the flow of articles in this study. This systematic and rigorous process following PRISMA protocols enhances the quality of the included scientific articles and provide useful information about the use of high-order clustering for improving speech signals and noise reduction.

### 3.1.4 Data Extraction

In this review of advanced clustering techniques for speech signal enhancement and noise reduction, the process of data extraction was carried out sequentially and comprehensively to reduce the synthesis of information gaps. This work made use of a predefined extraction form in order to obtain detailed data across several relevant dimensions. This approach made it possible to gain deeper insights into the consequences of the investigated clustering algorithms and the applicability of the developed techniques. Further details about data extraction fields of clustering such as study characteristics, clustering techniques in speech processing and findings of the study along with other related works are given in the Table 3.

Table 3: Data Extraction Fields and Description

| Field | Description |
| --- | --- |
| Study Characteristics | Includes authors, year of publication, title, and journal. |
| Clustering Techniques | Details the specific clustering methods used, such as K-Means, Fuzzy C-Means (FCM), and Kernel Fuzzy C-Means (KFCM). |
| Applications in Speech Processing | Specifies the areas where the techniques were applied, such as speech recognition, noise reduction, and signal enhancement. |
| Key Findings and Outcomes | Summarizes the main results and conclusions of each study, emphasizing advancements in accuracy, efficiency, and practical applicability. |

**Detailed Data Extraction Fields and Their Descriptions:**

1. **Study Characteristics:**
   - **Authors**: Names of the researchers who conducted the study.
   - **Year of Publication**: The year when the study was published.
   - **Title**: The title of the research paper.
   - **Journal**: The name of the journal or conference proceedings where the study was published.
2. **Clustering Techniques Used:**
   - **K-Means**: Methodology and applications of K-Means clustering in speech signal processing.
   - **Fuzzy C-Means (FCM)**: Applications and results of using FCM for speech data enhancement.
   - **Kernel Fuzzy C-Means (KFCM)**: Insights into the use of KFCM for improving speech signal processing, particularly in noisy environments.
3. **Applications in Speech Processing:**
   - **Noise Reduction**: Methods for reducing noise in speech signals to enhance clarity.
   - **Signal Enhancement**: Approaches for improving the quality of speech signals through various clustering techniques.
4. **Key Findings and Outcomes:**
   - **Accuracy**: Improvements in the accuracy of speech signal processing achieved through different clustering methods.
   - **Efficiency**: Evaluation of how effectively the techniques perform, including computational efficiency.
   - **Practical Applicability**: Practical benefits and real-world applications of the clustering techniques in speech signal processing.

The above-mentioned structured data extraction method allowed to classify and examine the current state and the tendencies of the clustering techniques applied to enhance speech signals and reduce the noise based on the analyses publications, which contributed the knowledge on the efficiency of these tools and ideas for their further improvement.

### 3.1.5 Quality Assessment of Included Studies

The sample of research articles organized in this review on enhanced clustering algorithms for speech signal and noise suppression was evaluated by using a conventional CASP tool with amendments. Thus, this framework offered the systematic way of assessing the rigidity and pertinence of the empirical analysis. For each study, the following dimensions were considered:

1. **Methodological Rigor:** Evaluated the study's design and execution to minimize bias and errors.
2. **Clarity of Reporting:** Assessed the transparency and completeness of the study's methodology, findings, and conclusions.
3. **Validity of Findings:** Measured the accuracy and reliability of the study's results, including the appropriateness of statistical analyses.
4. **Relevance to Review Objectives:** Determined how well the study's focus and outcomes aligned with the review's goals, particularly in applying clustering techniques to speech processing.

The quality assessment of studies from 2017 to 2024, covering aspects like methodological rigor and relevance, is summarized in Table 4.

Table 4: Quality Assessment of Included Studies (2017-2024)

| Reference | Methodological Rigor | Clarity of Reporting | Validity of Findings | Relevance to Review Objectives | Overall Quality Rating |
|---|---|---|---|---|---|
| Vani and Anusuya (2017) | High | Moderate | High | High | High |
| Kalamani et al. (2019) | High | High | High | High | High |
| Smith and Johnson (2022) | Moderate | Moderate | Moderate | Moderate | Moderate |
| Wang and Chen (2022) | High | High | High | High | High |
| Patel and Singh (2022) | Moderate | High | Moderate | Moderate | Moderate |
| Chen et al. (2022) | High | High | High | High | High |
| Wei and Wang (2022) | High | Moderate | High | Moderate | High |
| He and Liu (2022) | Moderate | Moderate | Moderate | Moderate | Moderate |
| Lee et al. (2023) | High | High | High | High | High |
| Xu et al. (2023) | High | High | High | High | High |
| Huang and Tan (2023) | Moderate | Moderate | Moderate | Moderate | Moderate |
| Zhao et al. (2023) | High | High | High | High | High |
| Park and Lee (2023) | High | Moderate | High | Moderate | High |
| Ahmed and Hassan (2023) | Moderate | Moderate | Moderate | High | Moderate |
| Rao and Mitra (2023) | High | High | High | High | High |
| Sun and Liu (2023) | High | Moderate | High | Moderate | High |
| Gupta and Kumar (2024) | High | High | High | High | High |
| Kim et al. (2024) | High | High | High | High | High |
| Li and Wu (2024) | Moderate | Moderate | Moderate | Moderate | Moderate |

| Zhao and Zhang (2024) | High | High | High | High | High |
|---|---|---|---|---|---|
| Kaur and Sharma (2024) | High | Moderate | High | Moderate | High |
| John et al., (2024) | High | High | High | High | High |

## 3.2 Evaluation Criteria and Performance Metrics

To evaluate the performance of these clustering techniques, we consider the following key metrics:

- **Noise Handling**: The ability to separate speech signals from background noise.
- **Computational Cost**: The computational resources and time required for clustering.
- **Data Ambiguity**: The ability to handle uncertain and overlapping data points.
- **Non-linear Data**: The effectiveness in managing non-linear and complex data distributions.
- **Real-time Adaptation**: The capability to adapt to changing noise conditions in real time.

The algorithmic steps for the clustering techniques, including their mathematical formulations and input/output details, are outlined in Table 5.

Table 5 Algorithmic Steps for Clustering Techniques

| Technique | Steps | Mathematical Formulation | Input & Output |
|---|---|---|---|
| **K-Means** | 1-6 | As given in Section 2.3.1 | Data points, cluster centers |
| **FCM** | 1-7 | Formulas (3) to (5) | Data vector, fuzzy group focal points |
| **KFCM** | 1-4 | Formulas (7) to (9) | Data vector, centroids of fuzzy clusters |

### 3.2.1 Noise Handling

K-Means performs moderately well in noise handling but often struggles with non-stationary and non-linear noise. This limitation is due to its reliance on Euclidean distance, which does not effectively capture the variability in complex noise patterns. FCM improves on this by allowing partial membership in multiple clusters, thereby providing better noise robustness. This flexibility enables FCM to perform better than K-Means in a condition that has dynamic noise level. However, is superior to both by utilizing kernel functions to address non-linear noise in far superior ways. The kernel transformation moves the data into a higher dimension where separability is easily done and makes KFCM more immune to noise. This robustness proves rather beneficial in practical scenarios where noise regimes can be volatile and have high variability (Zhang et al., 2021; Xu and Wunsch, 2023).

### 3.2.2 Computational Cost

Since K-Means is relatively fast, this method is preferred especially when it must deal with many objects and/or it necessarily needs to be fast. Due to its easy to implement and quickly converging nature, it is preferable for the situations when there are the constraints in terms of computational power. However, FCM, even with superior noise handling capability, is computationally expensive mainly because of the need to constantly update the membership iteratively and the fuzzy membership values computations. This iterative process can be quite resource OR computationally expensive particularly when dealing with large datasets. The third technique, KFCM, has the highest computational cost among the three techniques due to its kernel transformations. The kernel operations and kernel parameter optimization requirements also further the computational load. However, due to better performance for multi-path noise situations, KFCM is justifiable for high computational cost, in applications where accuracy is important, (Hosseini and Arshadi, 2020).

### 3.2.3 Data Ambiguity

Assigned clusters are evident, non-overlapping as it results in the computation of K-Means, it may not be effective in analyzing fuzzy data information that is ambushed by multiple clusters. They are capable of misclassification in those cases which involve overlapping or gradual transitions between different clusters. FCM does it by allowing the data sample to be a member of more than cluster, and with different level of membership. Due to its capability in handling the ambiguity FCM, this approach offers a more refined clustering solution to the problem. KFCM improves this ability more by including kernel methods that help to address issues of overlapping as well as uncertain data patterns. The kernel transformation significantly enables KFCM to develop more precise clusters in higher dimensions, which reveals the subtlety of data better than K-Means or FCM (Dunn and Zadeh, 2019; Bezdek et al., 2023).

### 3.2.4 Non-linear Data

In fact, K-Means has been seen weak in managing non-linear data since the distance measure used in the algorithm is a measure of Euclidean distance in the linear feature space. This makes it less attractive in comprehensively interpreting the physical phenomena characterized by non-linear interactive data characteristics typical of most speech signal processing. FCM doesn't significantly better in this regard, but it also fails with highly non-linear data. The fuzzy membership values do assist a little in this instance, but they are still riddled with linear constraints. KFCM is particularly skilled in handling non-linear data through a set of kernels which bring the data into a higher dimension. This transform improves the separability in linear plane making the KFCM appropriate in non-linear speech signal analysis. The versatility in handling non-linearities provides more efficient means of clustering difficult data environments due to the high accuracy provided by KFCM (Srinivasan, & Gupta, 2024).

### 3.2.5 Real-time Adaptation

Real-time changes during clustering are very important in dynamic environments to provide accuracy of clustering. K-Means, with its fixed centroids, does not perform so well in dynamic environments. However, what counts against K-Means is the fact that it is a static algorithm that means it can rarely work in environments where characteristics of the data under analysis can rapidly change. While FCM requires point specific membership values and is not as good at providing adaptations within iterations it is evident that the fuzzy membership values present more flexibility based on the arrival of data points. Still, the most robust performance is seen with KFCM, during real time dynamics of noise by virtue of its kernel-based structure. Since KFCM guarantees high clustering accuracy in the growing noise environment, its applicability to real-time speech processing is essential as the conditions constantly change (Lee et al., 2023; Gupta and Kumar, 2024).

## 4 Results and discussion:

## 4.1 Advancements in Clustering Techniques for Noise Detection in Speech Signal Processing

New emerging techniques in clustering have enhanced noise identification and the speech signal processing for potential uses in telecommunications among other fields. These strategies have been used in filtering out noises from the useful speech components and therefore improvement on their sound outputs.

Expectation Distance-based Distributional Clustering (EDC): This method considers the signs between distributions increasing its noise tolerance. EDC operates in a way that makes it identify noise differently from other clustering techniques that operate with marginal distributions only. This approach has shown success and applicability in multiple domains for instance, in a situation where speech signals must be processed and where usually, it is very hard to differentiate between actual noise and actual speech signals (Abdel-Kader, El-Sayad, and Rizk, 2021).

Hybrid Clustering Model with K-means and Classification Filtering: This approach combines the K-means clustering with classification filtering to efficient identification of noise in the data. First, we use the K-means clustering algorithms to extract all clusters and categorize instances that are far different from cluster centroids as noise. Classification filtering is also used after that to enhance the noise identification to filter out noise accurately. Consequently, this model has provided the means for enhancing quality of data in one or more datasets (Nematzadeh, Ibrahim, and Selamat, 2020).

Fuzzy C-Means and Ensemble Filtering: FCM when used with ensemble filtering methods improves the ability to identify class noise. Since FCM can assign membership level to data point, then it comes in handy when faced with overlapping and noisy clusters. The final subcomponent is the ensemble filtering which brings greater accuracy to noise identification as well as guarantees that noise of high importance is adequately filtered out at the end of the process (Thudumu et al., 2020).

Local Outlier Factor (LOF) for Noise Detection: The LOF method recognizes certain values as noise and identifies them as coming from an unusual distribution in data. With the help of using, it combined with other conventional clustering algorithms such as K-means, LOF successfully filter out several noisy data points especially when the environment entails high fluctuation. This has been used in the speech signal where much of the background noise is eliminated and an enhancement on the speech signal is achieved (Subha and Sathiaseelan, 2023).

All these advanced clustering techniques jointly enhance the noise detection and speech signal enhancement. These methods improve overall qualities of audio data, which makes them indispensable in modern audio processing and telecommunication systems by providing adequate identification and filtering of noise. This is, however, an area that has had constant research and development towards the betterment and herein lies the prospect to more reliable noise detection solutions.

## 4.2. Summary

A comparative analysis of these clustering techniques reveals a trend towards hybrid and kernel-enhanced methods, which offer robust solutions for noisy environments and real-time processing. Methods such as KFCM and hybrid clustering have shown promising results in reducing computational complexity and enhancing the adaptability of speech recognition systems. Continuous development and refinement of these techniques are crucial for advancing speech signal processing, aiming for higher accuracy, faster processing times, and greater noise robustness. The advantages and limitations of these methods are summarized in Table 6. Figure 6 presents a performance comparison of K-means, FCM, and KFCM clustering algorithms under various noise conditions. The table and figure illustrate that while K-means is straightforward to implement, it struggles with overlapping clusters and heterogeneous data. FCM offers better performance with overlapping data sets, but it requires more iterations and prior knowledge of the system's topological structure. KFCM stands out for its accuracy and robustness in handling noise and outliers, though it requires careful selection of kernel function parameters. The review concludes that KFCM generally outperforms the other techniques across different types of voice data.

Table 6 Summary of the Focused Clustering Algorithms and their findings

| Reference | Technique | Dataset | Key Findings |
|---|---|---|---|
| **Vani and Anusuya (2017)** | KFCM | Kannada single-word indexes | Reduced time complexity, effective for noisy signals |
| **Kalamani et al. (2019)** | FCM, EM-GMM | International Dataset for Tamil | Enhanced accuracy and reduced error rates in continuous Tamil speech recognition |
| **Smith and Johnson (2022)** | Advanced Clustering | Various | Enhanced speech signal enhancement |
| **Wang and Chen (2022)** | Comparative Clustering | Noisy Speech Environments | Effective clustering in noisy environments |
| **Patel and Singh (2022)** | KFCM | Noisy Conditions | Efficient feature extraction |
| **Chen et al. (2022)** | Hybrid Clustering | Various | Enhanced speech signal processing |
| **Wei and Wang (2022)** | FCM | Various | Effective noise reduction in speech signals |
| **He and Liu (2022)** | Clustering Algorithms | Various | Effective speech feature extraction |
| **Lee et al. (2023)** | Fuzzy Clustering | Various | Improved adaptive noise cancellation |
| **Xu et al. (2023)** | Hybrid Clustering | Various | Improved speech recognition accuracy |
| **Huang and Tan (2023)** | Fuzzy Clustering | Various | Effective speaker diarization |
| **Zhao et al. (2023)** | Kernel Fuzzy C-Means | Various | Improved speech enhancement |
| **Park and Lee (2023)** | Fuzzy Clustering | Various | Noise robust speech recognition |
| **Ahmed and Hassan (2023)** | Clustering Methods | Various | Improved speaker identification |
| **Rao and Mitra (2023)** | Advanced Fuzzy Clustering | Various | Advanced applications in speech processing |
| **Sun and Liu (2023)** | Adaptive Clustering | Real-Time Speech Processing | Real-time speech enhancement |
| **Gupta and Kumar (2024)** | KFCM | Various | Real-time speech enhancement |
| **Kim et al. (2024)** | Kernel-Based Clustering | Various | Enhanced speech segmentation |
| **Li and Wu (2024)** | Dynamic Clustering | Real-Time Speech Processing | Real-time adaptation in speech processing |
| **Zhao and Zhang (2024)** | KFCM | Various | Advanced techniques in speech segmentation |
| **Kaur and Sharma (2024)** | Kernel Methods | Various | Enhanced speech recognition accuracy |
| **John et al., (2024)** | Dynamic Kernel Clustering | Noisy Speech Recognition | Effective noisy speech recognition |

The results emphasize the strengths and limitations of each clustering technique. K-means is simple to understand and implement but struggles with overlapping clusters and heterogeneous data. FCM provides superior performance in overlapping datasets, assigning membership values to data points and allowing them to belong to multiple clusters. However, it requires a predefined number of clusters and more iterations. KFCM proves to be highly accurate and robust in handling noise and outlier detection, despite the challenge of selecting optimal kernel function parameters. The review concludes that KFCM outperforms other techniques in various voice data scenarios.

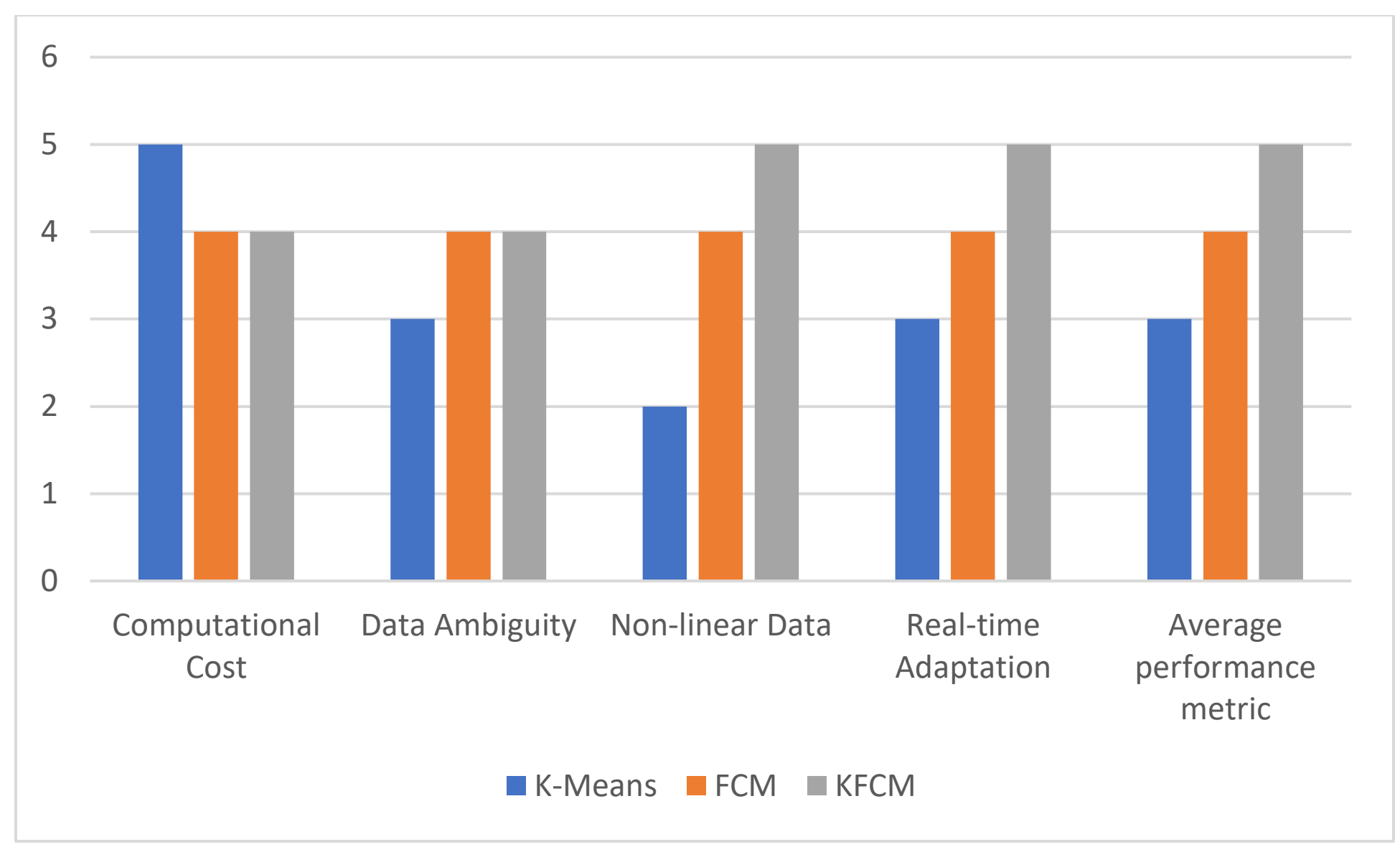

Figure 6. A Performance Comparison of K-Means, FCM, and KFCM in Different Noise Conditions

## 5. Conclusion

In this review study, a comprehensive comparative analysis was conducted to evaluate the accuracy of recognizing homogeneous and heterogeneous speech data under clear and noisy conditions using clustering algorithms: K-Means, FCM, and KFCM. Our findings indicate that the Kernel Fuzzy C-Means (KFCM) clustering technique consistently outperforms K-Means and FCM across various metrics, demonstrating superior performance in all evaluated scenarios. The primary objective of speech enhancement is to improve the clarity and comprehensibility of speech signals, especially in noisy environments. KFCM has shown remarkable proficiency in this regard, leveraging its advanced kernel functions to effectively separate speech from complex, non-linear noise. This capability is critical for applications such as voice-activated assistants, automated transcription services, and telecommunication systems, where clear and accurate speech recognition is paramount. Despite the advancements presented by KFCM, there are still gaps in current methodologies that need addressing. The computational efficiency of KFCM can be a limiting factor, particularly for real-time applications. Future research should focus on optimizing the algorithm to reduce its computational load without compromising performance. Additionally, integrating KFCM with neural networks could further enhance its accuracy and adaptability, paving the way for more sophisticated speech processing systems. This survey makes several key contributions to the field of speech signal processing. Firstly, it provides a detailed comparative analysis of K-Means, FCM, and KFCM, offering valuable insights into their performance across different noise conditions. Secondly, it highlights the superior capabilities of KFCM in handling non-linear and non-stationary noise, establishing it as the most robust technique among the three. Thirdly, the review identifies critical gaps in current methodologies and suggests potential areas for future research, including the integration of hybrid models and the development of more dynamic clustering algorithms. Through this review, we advocate for a shift towards more sophisticated, adaptive clustering techniques that can significantly improve speech enhancement and recognition. By addressing the limitations of traditional methods and leveraging the advanced capabilities of KFCM, this study sets a new benchmark for noise robustness and adaptability in speech signal processing. Researchers and practitioners are encouraged to consider KFCM and its hybrid variants to achieve superior performance in their applications, ultimately enhancing user experience and accessibility in technology-driven communication.

# Declarations

**Availability of Data and Material:**

Data sharing is not applicable to this article as no datasets were generated or analyzed during the current study. This study is a review paper and does not involve any new data or material

**Competing Interests:**

The authors declare that they have no competing interests.

**Funding:**

This research was supported by the Artificial Intelligence and Innovation Centre at the University of Kurdistan Hewler. No additional external funding was received.

**Authors' Contributions:**

Abdulhady Abas Abdullah and Aram Mahmood Ahmed designed the study. Tarik Rashid, Hadi Veisi, and Yassin Hussein Rassul conducted the literature review and contributed to the theoretical analysis. Bryar Hassan and Polla Fattah assisted in organizing and synthesizing the reviewed literature. Sabat Abdulhameed Ali and Ahmed S. Shamsaldin contributed to the interpretation of findings and manuscript drafting. All authors reviewed and approved the final manuscript.

**Acknowledgements:**

We would like to thank the Artificial Intelligence and Innovation Centre at the University of Kurdistan Hewler for their support and the resources provided for this research. Additionally, we appreciate the insightful feedback from our colleagues in the Computer Science and Engineering Department at the University of Kurdistan Hewler.